# A filtered embedded weighted compact nonlinear scheme for hyperbolic conservation law

Xuan Liu[1], Yaobing Min[2,*], Jinsheng Cai[1], Yankai Ma[2], Zhenguo Yan[2]

1 School of Aeronautics, Northwestern Polytechnical University, 710000, Xi'an, Shanxi

2 State Key Laboratory of Aerodynamics, 621000, Mianyang, Sichuan

**Abstract**

In situations where a wide range of flow scales are involved, the nonlinear scheme used should be capable of both shock capturing and low-dissipation. Most of the existing WCNS schemes are too dissipative because the weights deviate from ideal weights in the smooth regions caused by small-scale fluctuations. Moreover, due to the defect of the weighting strategy, the two smooth stencils located on the same side of a discontinuity cannot achieve fourth-order when the discontinuity only crosses S0 or S2. In this paper, we proposed the filtered embedded WCNS scheme which is applicable for complex flow simulations involving both shock and small-scale features. In order to overcome the above deficiency of existing WCNS scheme, a new mapping function is proposed to filter the weights deviation out which can map the weights to ideal weights in smooth region. Meanwhile, the embedded process also implemented by this function which is utilized to improve the resolution of shock capturing in certain discontinuity distributions. The approximate-dispersion-relation analysis indicates that the scheme with the mapping function we proposed has lower dispersion error and numerical dissipation as compared to the WCNS-JS and WCNS-Z schemes. The improved performance is demonstrated by the simulation of linear advection problem and nonlinear hyperbolic conservation laws.

# 1 Introduction

The simulations of high speed dynamics present a significant challenge due to the coexistence of small-scale features and discontinuities. The limiters or discontinuity sensors are used to add numerical dissipation near discontinuities and make the schemes have shock capturing capacity in traditional high-order nonlinear schemes. It is widely accepted that schemes that utilize fixed stencil invariably generate oscillations when discontinuities exist in the solution. The simplest of the nonlinear schemes is the first-order upwind scheme. Harten[1] applies the limiter to achieve second-order Total Variation Diminishing (TVD) which makes the method very popular in the simulation of supersonic flows. Although these schemes can solve the hyperbolic conservation law stably, the prohibitively dissipation and dispersion limits the application of these schemes.

To reduce the numerical errors, high-order accuracy numerical schemes are developed in the last two decades. Harten[3] proposed the essentially non-oscillatory (ENO) scheme, which divide five points into three candidate stencils and choose the smoothest stencil among them. However, the ENO scheme is only third-order in smooth region due to its strategy. To remedy the situation, Liu, et al[4] extend ENO and introduced weighted essentially non-oscillatory (WENO) scheme which weighted three candidate stencils in a convex combination and the contribution of stencil is determined by the smoothness of the solution over each stencil. The most applications of WENO scheme is designed by Jiang and Shu[5] which contains a different smoothness indicator and it is well known as WENO-JS. Nevertheless, WENO-JS scheme fails to achieve fifth-order when the first-order or higher-order derivations of the solution vanished at critical points.

Up to now, three different improved weighting strategies have been developed for reducing dissipation. The first strategy involves increasing the bias of the maximum-order upwind scheme in smooth regions. With the WENO-M[6] scheme proposed by Henrick et al, the weights are pushed smoothly to the optimal values by a mapping function. WENO-Z has a slightly different weighting formulation for it normalizes the smoothness indicators by a reference value which drives the weights to the optimal values faster than the classical WENO scheme. Another strategy introduces the downwind stencil which enables the scheme to use the optimal weight for central. WENO-SYMOO, WENO-SYMBO[9] and WENO-SYMBOO[10] schemes utilize the optimal central scheme for better bandwidth efficiency. With WENO-CU6[8], Hu et.al introduces downwind stencil and use the global smoothness indictor as the indictor of it which constructs a scheme that can switch between sixth-order and fifth-order scheme. The third strategy introduces a methodology to switch the scheme between the nonlinear and linear schemes, which can combine the low dissipation of linear scheme and stability of nonlinear scheme. Representatively, the targeted ENO(TENO) scheme introduced by Fu et al.[11] is applicable for the simulation involving a wide range of flow scales[12, 13]. However, due to its relatively direct switching strategy between linear and nonlinear scheme, the computational stability in the calculation needs to be considered.

Weighted compact nonlinear scheme (WCNS) combines the compact scheme and WENO scheme, leading to another high-order method for the discretization of hyperbolic conservation laws. Deng and Zhang[14] first proposed WCNS which combine the WENO interpolation and cell-centered compact schemes. Later Nonomura[15] considered the derivation scheme contains both solution points $f_i$ and flux points $f_{i+\frac{1}{2}}$ and compared to the scheme that only uses flux points, the WCNS-MND scheme has better performance in robustness. In addition, Nonomura[16] proposed the interpolation scheme is key to improve the resolution of WCNS. And hence, Zhang[17] and Nonomura[18] expand the fifth-order WCNS to ninth-order WCNS. Sumi and Kamiya[19] uses Hu et.al's strategy of WENO-CU6 and introduced WCNS-CU6. Wong[20] combined fifth-order WCNS and WCNS-CU6, the switch of two schemes are constructed by a shock sensor. Deng[21] constructed Dissipative Compact scheme (DCS) based on tridiagonal central compact stencils. Recently, based on the ideal of TENO, Toshihiko[22] constructed WCNS-T scheme which can capture strong discontinuities and high-frequency waves.

As mentioned above, we briefly reviewed the development of WENO and WCNS scheme. Recently, Bart et.al[23] noticed that the most of existing schemes discarded information of smooth stencils when only one of three candidate is non-smoothness. To improve the resolution near the discontinuity, the embedded WENO scheme is presented. The scheme improve the accuracy at transition points from discontinuous region to smooth region to fourth-order. Later Shen et.al[24] introduced a multistep WENO which is achieve by judgment. And Ma et.al[25] proposed a new smoothness indictor to construct embedded process. All of them are fourth-order when the stencil $S_0$ and $S_1$ contains discontinuity. In this paper, we consider the embedded/multistep method and expand it to WCNS framework. Then we introduced a new embedded WCNS/WENO scheme with filter property by a different approach. It filers out the weight perturbations near the control weights and improves the resolution of the scheme.

The rest of the paper is organized as follows: Section 2 briefly present the classical fifth-order WCNS scheme and the construction of frequently-used weights. In section 3, we propose the process of constructing fifth-order Filtered Embedded WCNS scheme(WCNS-FE). The numerical results are show in section 4. Section 5 summarizes the work. In addition, we show the statistical probability of WCNS-Z and WCNS-FE in Appendix. A. and the extension of WENO is shown in Appendix. B.

## 2 The classical WCNS scheme

In this section, we review fifth-order WCNS-JS, WCNS-M, and WCNS-Z scheme, which is described by the hyperbolic system of conservation laws,

$$\frac{\partial u}{\partial t} + \frac{\partial f(u)}{\partial x} = 0, \tag{1}$$

where, $u = u(x,t)$ is the conservative variable, $f(u)$ stands for the flux. The computational domain is divided into $N$ uniform cells with spatial step $h = \frac{1}{N}$ which is staggered by solution points and flux points. And $x_j = (j-1)h$ are solution points and $x_{j+\frac{1}{2}} = (j-\frac{1}{2})h$ are flux points.

Eq.(1) can be written as an ordinary differential equations.

$$\frac{d\hat{u}_i}{dt} = -\left.\frac{\partial f}{\partial x}\right|_{x=x_i} \tag{2}$$

The flux derivative at the solution point is obtained by sixth-order difference scheme,

$$\left.\frac{\partial f}{\partial x}\right|_{x=x_i} = \frac{d_1}{h}\left(f_{i+\frac{1}{2}} - f_{i-\frac{1}{2}}\right) + \frac{d_2}{h}\left(f_{i+\frac{3}{2}} - f_{i-\frac{3}{2}}\right) + \frac{d_3}{h}\left(f_{i+\frac{5}{2}} - f_{i-\frac{5}{2}}\right), \tag{3}$$

where, $d_1 = \frac{75}{64}, d_2 = \frac{-25}{384}, d_3 = \frac{3}{640}$.

The flux point values are derived from the solution point values found on a stencil. The flux function comes from a (2r-1)th-order Lagrange interpolation of the flux present on the stencil $\{x_{i-r+1},...,x_{i+r-1}\}$. We can evaluated the variables at any interface with the Lagrange interpolation polynomial.

$$\hat{u}_{i+\frac{1}{2}} = u_i + \sum_{k=1}^{2r-2} a_k (x - x_i)^k \tag{4}$$

The fifth-order WCNS interpolation uses three stencils,

$$S_0 = \{I_{i-2}, I_{i-1}, I_i\}, S_1 = \{I_{i-1}, I_i, I_{i+1}\}, S_2 = \{I_i, I_{i+1}, I_{i+2}\}$$

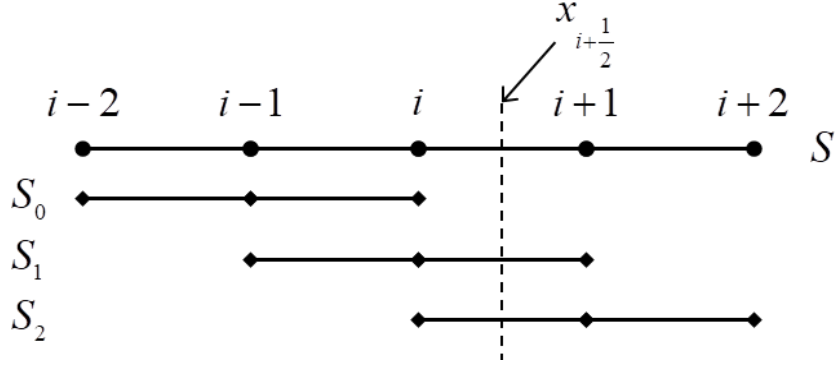

Fig. 1. The five point stencil, composed of candidate three-points stencils $S_0, S_1$, and $S_2$

We present the result of sub-stencil interpolation.

$$\hat{u}_{i+\frac{1}{2},0} = \frac{1}{8}(3u_{i-2} - 10u_{i-1} + 15u_i)$$

$$\hat{u}_{i+\frac{1}{2},1} = \frac{1}{8}(-u_{i-1} + 6u_i + 3u_{i+1}) \tag{5}$$

$$\hat{u}_{i+\frac{1}{2},2} = \frac{1}{8}(3u_i + 6u_{i+1} - \hat{u}_{i+2})$$

The grid point's flux calculation relies on the approximate Riemann solution.

$$u_{i+\frac{1}{2}} = \sum_{k=0}^{r-1} \omega_{i,k} u_{i+\frac{1}{2},k} \left( \hat{u}^L_{i+\frac{1}{2},k}, \hat{u}^R_{i+\frac{1}{2},k} \right) \tag{6}$$

For smooth region in the fluid domain, we denote ideal weights as $d_l$ $d_0 = \frac{1}{16}, d_1 = \frac{10}{16}, d_2 = \frac{5}{16}$ and $\omega_{j,k}$ is defined as follows.

$$\omega_{i,k} = \frac{\alpha_{i,k}}{\sum_{l=0}^{r-1} \alpha_{i,k}} \tag{7}$$

$$\alpha_{i,k} = \frac{d_k}{IS_{i,k}^2 + \varepsilon} \tag{8}$$

$$IS_{i,k} = (hf_{i,k})^2 + (h^2 s_{i,k})^2 \tag{9}$$

$$\begin{cases} f_{i,0} = \frac{1}{2h}(u_{i-2} - 4u_{i-1} + 3u_i), & s_{i,0} = \frac{1}{h^2}(u_{i-2} - 2u_{i-1} + u_i) \\ f_{i,1} = \frac{1}{2h}(u_{i+1} - u_{i-1}), & s_{i,1} = \frac{1}{h^2}(u_{i-1} - 2u_i + u_{i+1}) \\ f_{i,2} = \frac{1}{2h}(-3u_i + 4u_{i+1} - u_{i+2}), & s_{i,2} = \frac{1}{h^2}(u_i - 2u_{i+1} + u_{i+2}) \end{cases} \tag{10}$$

And we briefly present the calculation process of JS weights[5], M weights[6], and Z weights[7].

➢ JS weights

$$\omega_{i,k}^{JS} = \frac{\alpha_{i,k}^{JS}}{\sum_{l=1}^{r-1}\alpha_{i,l}^{JS}}, \alpha_{i,l}^{JS} = \frac{d_l}{(IS_{i,l}+\varepsilon)^2}, k=0,...,r-1 \tag{11}$$

➢ M weights

The scheme uses JS weights will lose accuracy at critical points, M weights construct a mapping function of the nonlinear weights $\omega$ are given by

$$\omega_{i,k}^{M} = \frac{\alpha_{i,k}^{M}}{\sum_{l=1}^{r-1}\alpha_{i,l}^{M}}, \alpha_{i,k}^{M} = g_{i,k}^{M}(\omega_{i,k}^{JS}) \tag{12}$$

$$g_{i,k}^{M}(\omega) = \frac{\omega_{i,k}(d_k + d_k^2 - 3d_k\omega_{i,k} + w_{i,k}^2)}{d_k^2 + (1-2d_k)w_{i,k}}, w_{i,k} \in [0,1], k=0,...,r-1 \tag{13}$$

➢ Z weights

Z weights has the same purpose as M weights, the unnormalized and normalized nonlinear weights $\alpha_k^Z$ and $w_k^Z$ are respectively defined by

$$\omega_{i,k}^{Z} = \frac{\alpha_{i,k}^{Z}}{\sum_{l=0}^{r-1}\alpha_{i,l}^{Z}}, k=0,...,r-1, \ \alpha_{i,k}^{Z} = d_k[1+(\frac{\tau}{(IS_{i,k}+\varepsilon)})^p] \tag{14}$$

If $r=3$, that is, when the interpolation accuracy is fifth-order.

$$\tau = |IS_{i,0} - IS_{i,2}| \tag{15}$$

Where $\varepsilon$ is a small number which is added to avoid division by zero.

In the end, we can solve the ordinary differential system by 4th Runge-Kutta method. This method is given by

$$\begin{cases} u^1 = u^n + \frac{\Delta t}{2}L(u^n) \\ u^2 = u^n + \frac{\Delta t}{2}L(u^1) \\ u^3 = u^n + \Delta t L(u^2) \\ u^{n+1} = -\frac{1}{3}u^n + \frac{1}{3}u^1 + \frac{2}{3}u^2 + \frac{1}{3}u^3 + \frac{\Delta t}{6}L(u^3) \end{cases} \tag{16}$$

## 3 Filtered Embedded WCNS scheme

In this section, we propose the concept of the filtered embedded WCNS scheme，and also the optimization of free parameter by ADR analysis is proposed.

### 3.1 Concept of WCNS FE scheme

We can get fifth-order scheme in smooth region and third-order result when discontinuity exist in the stencil by classical WCNS/WENO. However, if a discontinuity exists in the stencil, dividing the three sub-stencils into two smooth adjacent sub-stencils and one discontinuous sub-stencil, the optimal accuracy should be fourth-order. The classical WCNS/WENO scheme, however, can only achieve third-order due to an inapposite weighting principle. It is obvious that the scheme can't have optimal accuracy at transition points from discontinue region to smooth region due to the scheme generates a linear combination of the two smooth stencils by the principal of fifth-order construction.

The embedded/multistep nonlinear schemes overcome the disadvantage of the classical WCNS/WENO by pairing adjacent smooth sub-stencils and revising the calculation process of weights to create a larger sub-stencil that is fourth-order.

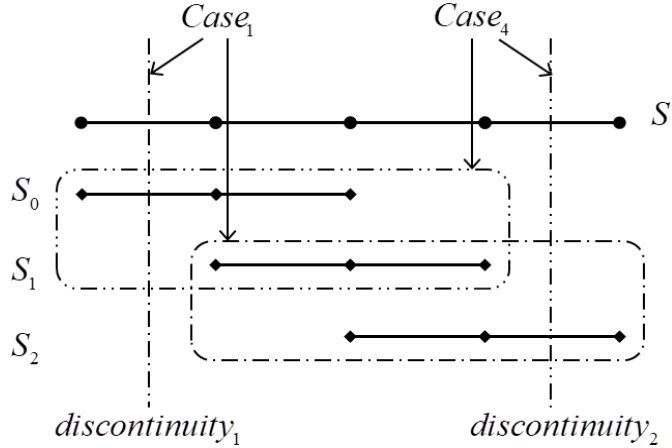

Fig. 2. Embedded scheme constructing process

For a fifth-order scheme, discontinuity may occur in four locations $\{[x_{i-2}, x_{i-1}], [x_{i-1}, x_i], [x_i, x_{i+1}], [x_{i+1}, x_{i+2}]\}$, these cases are successively named $Case_1 \sim Case_4$ and we call it $Case_0$ when all sub-stencils are smooth.

When the discontinuity crosses the sub-stencil, the weight will be 0 or a number approaching 0. So we can obtain the original weights $\omega_i^{original}$ of different cases.

$$\omega_i^{original} = \frac{\omega_i}{\sum_{j=0}^{2} \omega_j}, \quad i = 0, 1, 2 \tag{17}$$

$Case_1$ and $Case_4$ have two smooth adjacent sub-stencils which can be combined into a larger stencil. We can get the mapped weights by Taylor expand since the optimal accuracy of larger stencil is fourth-order.

As mentioned above, the truncation error of different cases are showed as follows.

$$\begin{aligned}
Trunction_{Case_0} &= \frac{3h^5 \left(\frac{d^5}{dx_0^5} f(x_0)\right)}{256} + \frac{h^6 \left(\frac{d^6}{dx_0^6} f(x_0)\right)}{1024} \\
Trunction_{Case_1} &= \frac{3h^4 \left(\frac{d^4}{dx_0^4} f(x_0)\right)}{128} + \frac{3h^5 \left(\frac{d^5}{dx_0^5} f(x_0)\right)}{256} \\
Trunction_{Case_2} &= \frac{h^3 \left(\frac{d^3}{dx_0^3} f(x_0)\right)}{16} + \frac{7h^4 \left(\frac{d^4}{dx_0^4} f(x_0)\right)}{128} \\
Trunction_{Case_3} &= \frac{5h^3 \left(\frac{d^3}{dx_0^3} f(x_0)\right)}{16} - \frac{25h^4 \left(\frac{d^4}{dx_0^4} f(x_0)\right)}{128} \\
Trunction_{Case_4} &= -\frac{5h^4 \left(\frac{d^4}{dx_0^4} f(x_0)\right)}{128} + \frac{3h^5 \left(\frac{d^5}{dx_0^5} f(x_0)\right)}{256}
\end{aligned} \tag{18}$$

We can get the correspondence between the original value and mapped value. And the function should equal to the mapped value at original value so that the scheme could be fourth-order in $Case_1$ and $Case_4$.

The embedded mapped weights $\omega_i^{mapped}$ can be introduced as follows.

- If the region is smooth, then $\beta_0 \cong \beta_1 \cong \beta_2$. According to Equation (17) and (18), we have $\omega_0^{original} = \omega_0^{mapped} \to 1/16, \omega_1^{original} = \omega_1^{mapped} \to 10/16, \omega_2^{original} = \omega_2^{mapped} \to 5/16$. And the numerical flux accuracy remains fifth-order.

- If the discontinuity located at $\{x_{i-2}, x_{i-1}\}$, then $\beta_0 \gg \beta_1, \beta_2$. According to Equation (17) and (18), we have $\omega_0^{original} = \omega_0^{mapped} \to 0$, $\omega_1^{original} \to 2/3, \omega_1^{mapped} \to 1/2$ and $\omega_2^{original} \to 1/3, \omega_2^{mapped} \to 1/2$. The numerical flux accuracy remains fourth-order.

- If the discontinuity located at $\{x_{i-1}, x_i\}$, then $\beta_0, \beta_1 \gg \beta_2$. According to Equation (17) and (18), we have $\omega_0^{original} = \omega_0^{mapped} \to 0$, $\omega_1^{original} = \omega_1^{mapped} \to 0$ and $\omega_2^{original} = \omega_2^{mapped} \to 1$. The numerical flux accuracy remains third-order.

- If the discontinuity located at $\{x_i, x_{i+1}\}$, then $\beta_1, \beta_2 \gg \beta_0$. According to Equation (17) and (18), we have $\omega_0^{original} = \omega_0^{mapped} \to 1$, $\omega_1^{original} = \omega_1^{mapped} \to 0$ and $\omega_2^{original} = \omega_2^{mapped} \to 0$. The numerical flux accuracy remains third-order.

- If the discontinuity located at $\{x_{i+1}, x_{i+2}\}$, then $\beta_2 \gg \beta_0, \beta_1$. According to Equation (17) and (18), we have $\omega_0^{original} \to 1/11, \omega_0^{mapped} \to 1/6, \omega_1^{original} \to 10/11, \omega_1^{mapped} \to 5/6$ and $\omega_2^{original} = \omega_2^{mapped} \to 0$. The numerical flux accuracy remains fourth-order.

The result of embedded interpolation process can be expressed as Table 1 briefly.

Table 1. The mapping correspondence for embedded fifth order WCNS schemes

| | $\omega_0$ | | $\omega_1$ | | $\omega_2$ | |
|---|---|---|---|---|---|---|
| | $\omega_0^{original}$ | $\omega_0^{mapped}$ | $\omega_1^{original}$ | $\omega_1^{mapped}$ | $\omega_2^{original}$ | $\omega_2^{mapped}$ |
| Case$_0$ | $\dfrac{1}{16}$ | $\dfrac{1}{16}$ | $\dfrac{10}{16}$ | $\dfrac{10}{16}$ | $\dfrac{5}{16}$ | $\dfrac{5}{16}$ |
| Case$_1$ | - | - | $\dfrac{2}{3}$ | $\dfrac{1}{2}$ | $\dfrac{1}{3}$ | $\dfrac{1}{2}$ |
| Case$_2$ | - | - | - | - | 1 | 1 |
| Case$_3$ | 1 | 1 | - | - | - | - |
| Case$_4$ | $\dfrac{1}{11}$ | $\dfrac{1}{6}$ | $\dfrac{10}{11}$ | $\dfrac{5}{6}$ | - | - |

In order to avoid conceptual confusion, we denote the original weights $\omega_i^{original}$ and mapped weights $\omega_i^{mapped}$ shown in Table 1 as control point and control value of mapping function.

The above statement is one of the purposes of our schemes. Except for this, it is important to make the mapping function equal or approach to the control value at and near the control point which can filer out the weight perturbations caused by the high wavenumber structure appearing in turbulence simulation.

In order to prevent excessive destruction of the stability of the original scheme, the mapped weight should be recovered to the original weight when it exceeds the mapping control threshold. We design the following mapping function through the hyperbolic tangent function which is written as tanh(x) as well. The hyperbolic function is written as follows.

$$\tanh(x) = \frac{\sinh(x)}{\cosh(x)} = \frac{e^x - e^{-x}}{e^x + e^{-x}} \tag{19}$$

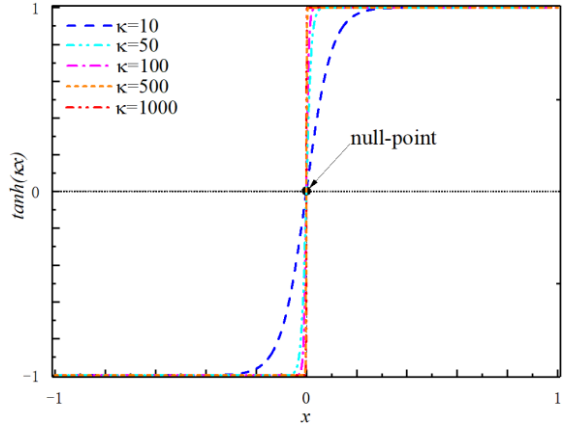

Fig. 3. $\tanh(\kappa x)$ variations of $\kappa$

The mapping function can be constructed by shifting the hyperbolic tangent function and multiple linear functions. To simplify the expression, we write the mapping function as follows.

$$\omega_i^{mapped} = g(\omega_i) = \sum_{j=1}^{4} \phi(i,j) \frac{\omega_{i,j} - \omega_{i,j-1}}{2} \tanh(\xi \cdot \phi(i,j) \cdot (\omega_i - \kappa(\psi_{i,j} + \psi_{i,j-1}))) + \omega_i$$

(20)

$$\phi(i,j) = \begin{cases} -1, (i,j) = \{(0,3) \cup (2,3) \cup (1,4)\}, \\ 1, else. \end{cases} \tag{21}$$

$\omega_i^{mapped}$ is the FE weights we proposed, $\frac{\omega_{i,j} - \omega_{i,j-1}}{2}$ is the amplitude control part which is determined by the control value. $\kappa(\psi_{i,j} + \psi_{i,j-1})$ is the position of hyperbolic function's null-point, which can show the distance of the function moves along X-axis and it is determined by adjacent control points. $\omega_i$ is the distance moves along Y-axis. $\xi$ is the coefficient that determines the steepness of the function near null point.

Table 1. The coefficient of mapping function

|  | i=0 | | i=1 | | i=2 | |
| --- | --- | --- | --- | --- | --- | --- |
|  | $\omega_{0,j}$ | $\psi_{0,j}$ | $\omega_{1,j}$ | $\psi_{1,j}$ | $\omega_{2,j}$ | $\psi_{2,j}$ |
| j=0 | $\omega_0$ | 0 | $\omega_1$ | 0 | $\omega_2$ | 0 |
| j=1 | $\frac{1}{16}$ | $\frac{1}{16}$ | $\frac{10}{16}$ | $\frac{10}{16}$ | $\frac{5}{16}$ | $\frac{5}{16}$ |
| j=2 | $\frac{1}{6}$ | $\frac{1}{11}$ | $\frac{1}{2}$ | $\frac{2}{3}$ | $\frac{1}{2}$ | $\frac{1}{3}$ |
| j=3 | $\omega_0$ | 1 | $\frac{5}{6}$ | $\frac{10}{11}$ | $\omega_2$ | 1 |
| j=4 | - | - | $\omega_1$ | 1 | - | - |

$\xi$ is the parameter to control the steepness of the function near the null-points, and $\kappa$ is set to move the position of null-points. For simplicity, $\xi$ is denoted as steepness parameter and $\kappa$ as null-points parameter. Both of them are free parameters that we will discuss next.

As the massive computing costs the mapping function is needed, we set an activation switch of mapping function as shown in Appendix A, and named it PFE weight, which is a variation of FE weight. The FE weight and PFE weight together constitute the FE family weight. As the calculation results are almost identical, we will use FE weights to illustrate in this paper.

**3.2 Unified framework for spectral property optimization**

In this section, the spectral property optimization is presented by the mathematic formulations. The dispersion and dissipation optimization of fifth-order WCNS scheme is given in detail. The property of WCNS-FE scheme is studied by ADR[26].

Firstly, we give the mapping function of $\omega_0^{mapped}$ varies by $\xi$ and $\kappa$. The other weights are similar and omitted here for brevity.

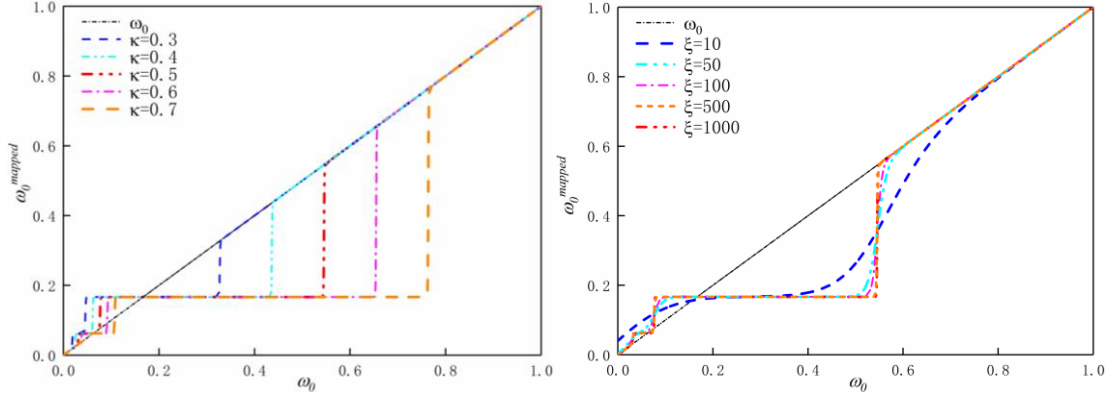

Fig. 4. Mapping function variations of $\xi$ and $\kappa$

The one-dimensional scalar conservation law is ta006en into consideration.

$$\frac{\partial u}{\partial t} + c\frac{\partial u}{\partial x} = 0, -\infty < x < +\infty \tag{22}$$

where the advection velocity $c$ is constant. A system of ordinary differential equation is shown below.

$$\frac{du_j}{dt} = -cu'_j, j = 0...N \tag{23}$$

the first spatial derivate $u'_j$ can be approximated by

$$u'_j = \frac{1}{h}\sum_{l=-r}^{r}\omega_l u_{j+l} \tag{24}$$

$\omega_l$ is the nonlinear coefficient depending on specific scheme.

Assuming the initial condition is sine function with wavelength $\lambda$ and wavenumber $\omega = \frac{2\pi}{\lambda}$.

$$Q(x,0) = Q_0 e^{i\omega x} \tag{25}$$

where $Q_0$ is initial amplitude. The analytical derivation of initial wave is given by

$$Q'_{Analytical}\Delta x = k_{Analytical}Q_0 e^{i\omega x} = (k_{r,Analytical} + ik_{i,Analytical})Q_0 e^{i\omega x} \tag{26}$$

$$\begin{cases} k_{r,Analytical} = 0, \\ k_{i,Analytical} = \alpha. \end{cases} \tag{27}$$

$k_{r,Analytical}$ and $k_{i,Analytical}$ are the real part and imaginary part of the analytical wave, $\alpha = \omega\Delta x$. Then we can obtain the numerical derivation of initial wave.

$$Q'_{Numerical}\Delta x = k_{Numerical}Q_0 e^{i\omega x} = (k_{r,Numerical} + ik_{i,Numerical})Q_0 e^{i\omega x} \qquad (28)$$

With this framework, the variation of dispersion and dissipation property with $\xi$ and $\kappa$ are shown in Fig.5 and Fig.6.

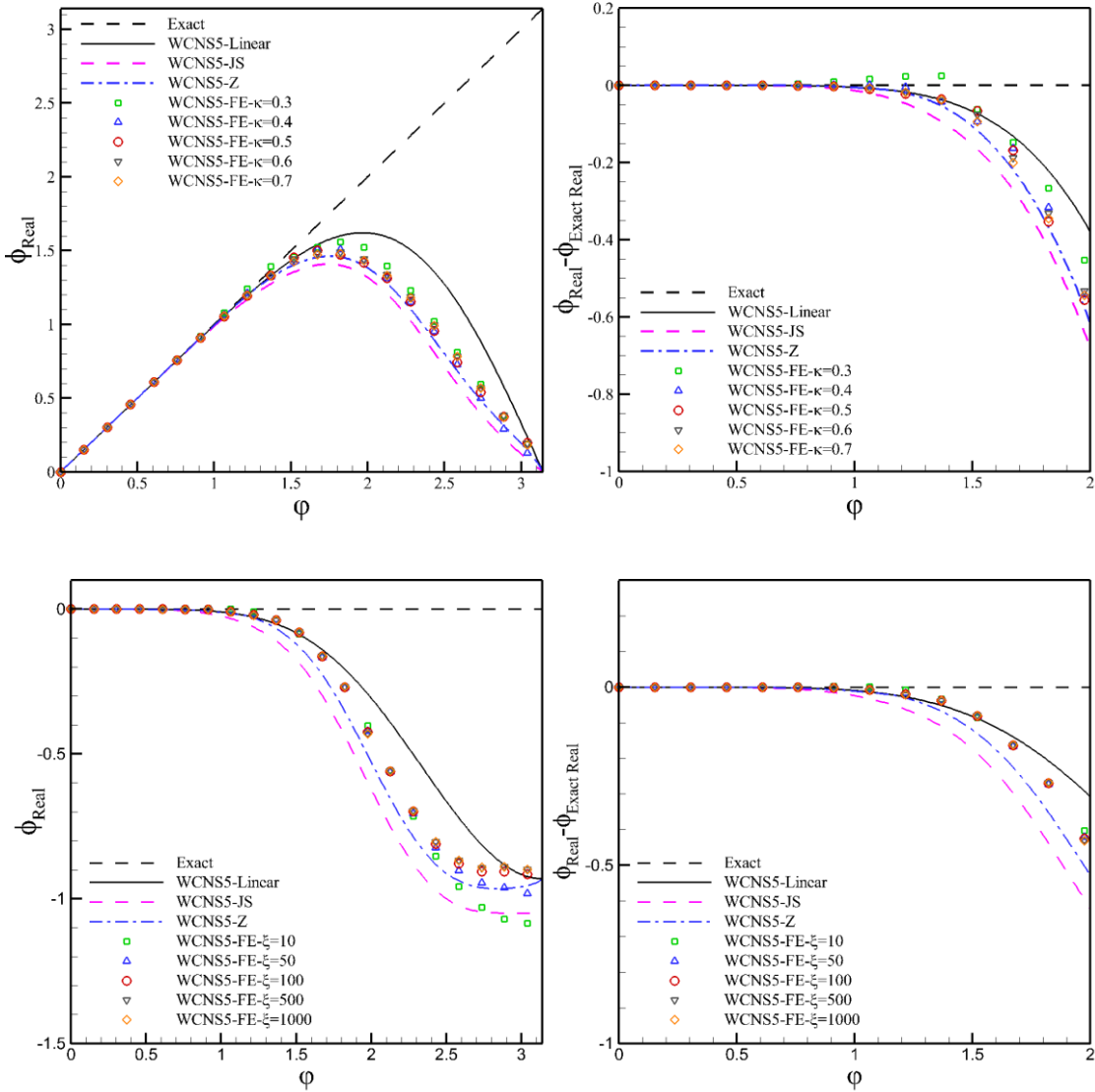

Fig. 5. The approximate dispersion relation for fifth-order weighted compact nonlinear scheme (WCNS) variation of $\kappa$ ($\xi = 10^6$)

In order to achieve a sharper junction area, $\xi$ should be large enough to eliminate its effect, And hence, $\xi = 10^6$ is the value taken here. The change of the approximate dispersion relation with $\kappa$ is shown in Fig.5. which reveals that there is an overshoot in the dispersion relations at $\kappa=0.3$ and $\kappa=0.4$, which can be attributed to the mapping function not reaching the control value at the control points due to an unreasonable value of $\kappa$. As such, it is recommended that the null-point parameter $\kappa$ be increased to a value greater than 0.5, and specifically, $\kappa=0.5$ is suggested to minimize dissipation and dispersion.

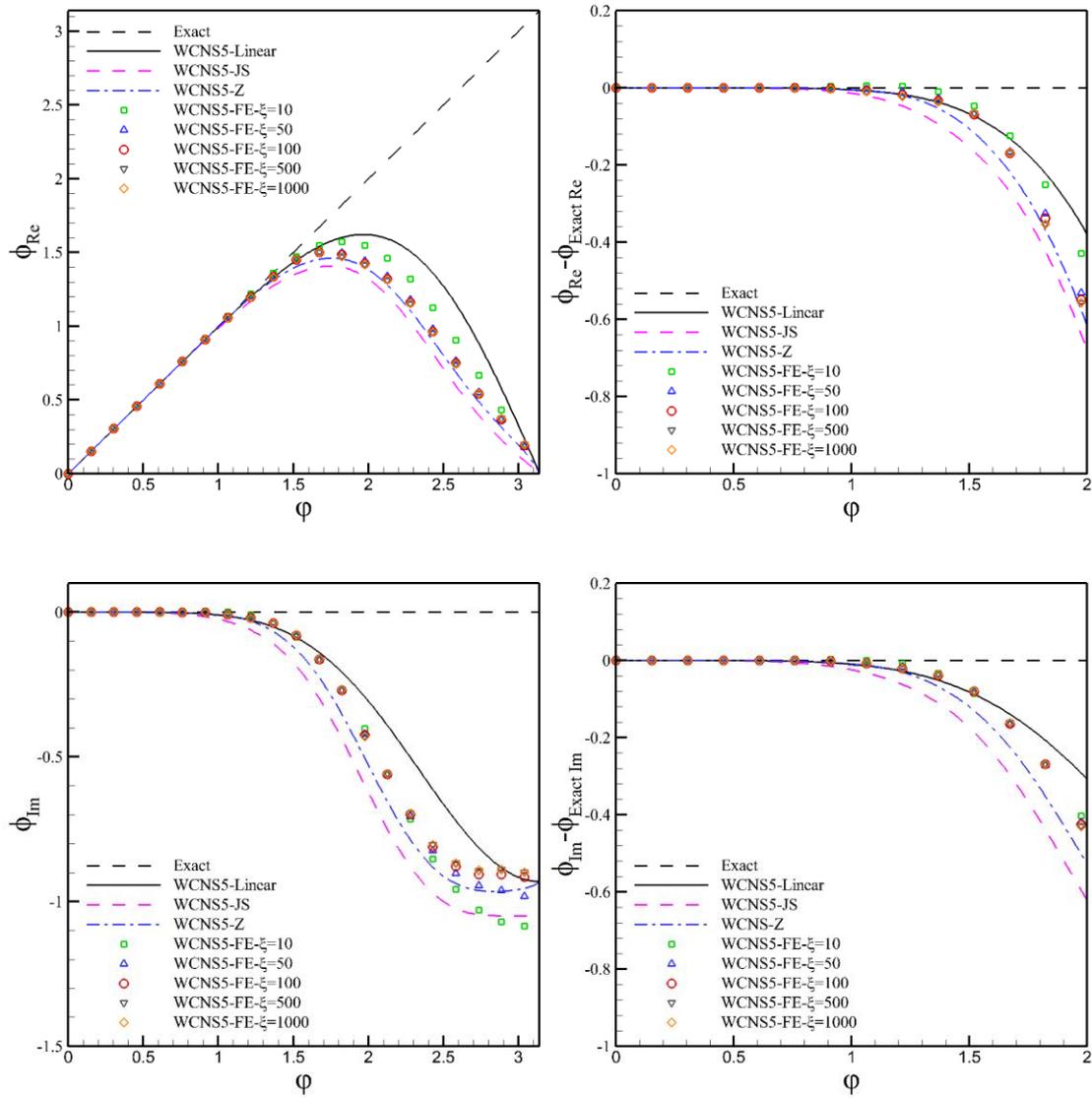

Fig. 6. The approximate dispersion relation for fifth-order weighted compact nonlinear scheme (WCNS) variation of $\xi$ ($\kappa=0.5$)

Take a closer look at Fig.6, when $\xi=10$, overshoot exists in the approximate dispersion relations. And hence, it is an appropriate value for the steepness parameter $\xi$ greater than 10. Furthermore, the approximate dispersion relation is nearly identical when $\xi$ is greater than 10. In fact, the choice of $\xi$ is subjective, $\xi=100$ is selected in this paper to ensure that the function is sharp enough in the junction area.

## 4 Numerical Experiments

To assess the performance of WCNS-FE, we compare it with other fifth-order WCNS schemes, such as WCNS-JS, WCNS-Z, WCNS-T, and WCNS-MYK [23]. The aim of this section is to test shock, low wavenumber and high wavenumber problems. Seven classical problems are tested, including the one-dimensional linear advection equation, 1D and 2D Euler systems. The numerical update is marched by fourth-order Runge-Kutta scheme with a CFL number set to be 0.3 and a fixed $\varepsilon = 10^{-40}$ for all schemes.

### 4.1 One-dimensional linear advection equation

In this subsection, one-dimensional advection equation has been tested.

$$u|_{t=0} = \begin{cases} \frac{1}{6}[G(x,z-\delta)+4G(x,z)+G(x,z+\delta)], & x \in [-0.8,-0.6], \\ 1, & x \in [-0.4,-0.2], \\ 1-|10x-1|, & x \in [0.0,0.2], \\ \frac{1}{6}[F(x,a-\delta)+4F(x,a)+F(x,a+\delta)], & x \in [0.4,0.6], \\ 0, & else. \end{cases} \quad (29)$$

where,

$$G(x,z) = e^{-\beta(x-z)^2}, F(x,z) = \sqrt{\max(1-\alpha^2(x-a)^2,0)}. \quad (30)$$

Periodic boundary condition is applied here, and the final time $t = 6.0$

Take a closer look at Fig. 7, our scheme is better than the other methodology shown below which are not mapped by our function at corner of the shock apparently.

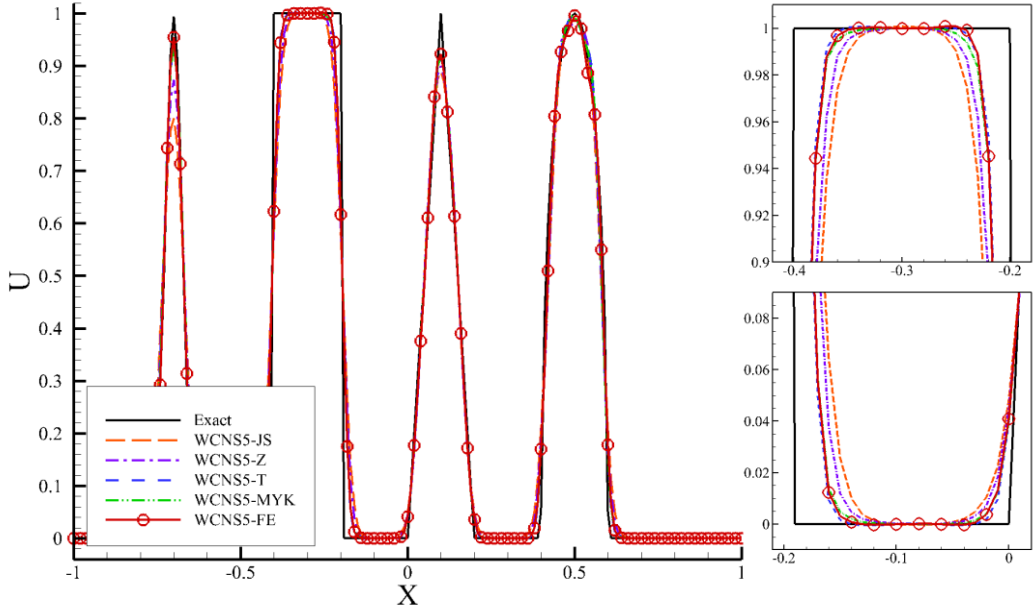

Fig. 7. Numerical solutions of the advection equation with the discontinuous initial condition Equation(30) where $\varepsilon = 10^{-40}$ as computed by fifth-order weighted compact nonlinear schemes(WCNS) with $N$=201 at $t$=6.0

### 4.2 One-dimensional Euler systems

Numerical experiment with the one dimensional Euler equations in conservation form are presented here.

$$\frac{\partial \mathbf{U}}{\partial t} + \frac{\partial \mathbf{F}}{\partial x} = 0 \tag{31}$$

where,

$$\begin{aligned} \mathbf{U} &= (\rho, \rho u, E)^T \\ \mathbf{F} &= (\rho u, \rho u^2 + p, (E+p)u)^T \end{aligned} \tag{32}$$

The density, velocity, pressure and total energy is denoted as $\rho, u, p$ and $E$. The shock tube and Shu-Osher problems are solved using 200 grids, with the exception of the reference result which used 2000 grids. Meanwhile, the Titarev-Toro problem is solved using 2000 grids and its reference result is obtained using 20000 grids.

### 4.2.1 Shock tube problem

This is a one-dimensional shock tube problem introduced by Sod[26] and Lax[27]. Both Sod and Lax problem consists of the propagation of a shock wave, a contact discontinuity and an expansion fan. The initial conditions are shown as follows,

$$\text{Sod}:(\rho,u,p) = \begin{cases} (1.0, 0, 1.0), 0 \leq x \leq 5, \\ (0.125, 0, 0.1), 5 < x \leq 10. \end{cases} \quad (33)$$

$$\text{Lax}:(\rho,u,p) = \begin{cases} (0.445, 0.698, 3.528), 0 \leq x \leq 0.5, \\ (0.5, 0, 0.571), 0.5 < x \leq 1. \end{cases} \quad (34)$$

the final time is $t = 2.0$ and $t = 0.15$.

Fig. 8 and Fig. 9 present the simulation results of Sod problem and Lax problem. As depicted, all schemes are able to capture the contact discontinuity without any spurious oscillations. However, effective shock capturing is only achieved by WCNS5_T and WCNS5_FE, possibly due to their superior filter property ensuring more adequate weight distribution when compared to WCNS5_JS, WCNS5_Z, and WCNS5_MYK, as indicated in Table A.1.

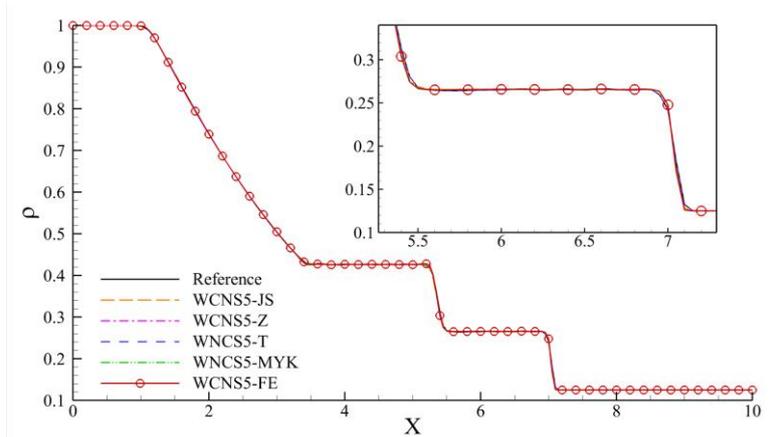

Fig. 8. Density profiles of the Sod problem computed by fifth-order weighted compact nonlinear schemes(WCNS) with $N$=201 points.

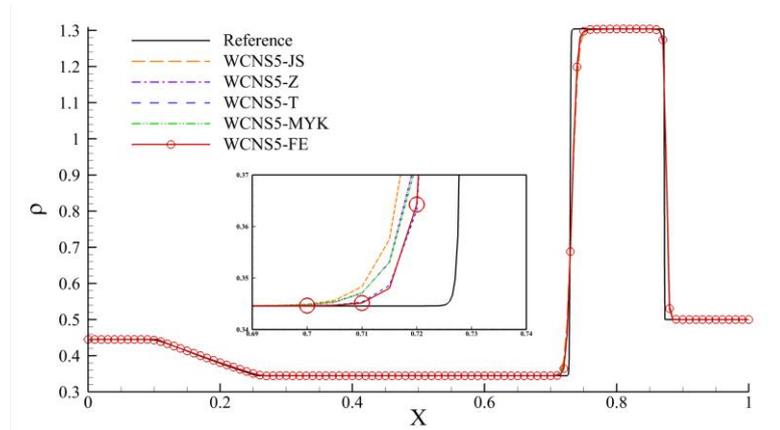

Fig. 9. Density profiles of the Lax problem computed by fifth-order weighted compact nonlinear schemes(WCNS) with $N$=201 points.

### 4.2.2 Modified shock/turbulence interaction

These problems describe the interaction of right-traveling shock and entropy waves and we solve the Euler equations with the following initial condition defined on $[0,10]$.

$$Shu-Osher:(\rho, u, p)=\begin{cases}(3.857143, 2.629369, 10.3333), x \in [0,1], \\ (1+0.2\sin(5x), 0, 1), x \in (1, 10].\end{cases} \quad (35)$$

$$Titarev-Toro:(\rho, u, p)=\begin{cases}(1.515695, 0.523346, 1.805), x \in [0, 3.5], \\ (1+0.1\sin(20\pi x), 0, 1), x \in (3.5, 10].\end{cases} \quad (36)$$

High wavenumber problems is described by Shu-Osher[5] and Titarev-Toro[28] cases. They both consist of a right-facing shock impinging into a high-frequency density perturbation, which can test the scale-separation capability of different schemes in capturing discontinuities and resolving smooth fluctuating waves. The computation marches up to t=1.8 and 4.0 with 201 and 2001 grids. We show the density profile enlargements of the solutions which contain the shock wave and smooth part. As we can see that all of the schemes can capture the shock wave without spurious oscillations. The WCNS-FE resolves the fluctuating best, even better than WCNS-T in both cases.

As we can see in Appendix A.2, the distribution of Z weights near the ideal weights is pretty small due to the high wavenumber cause the weight oscillate violently which identify high wavenumber region as discontinuity. The FE weights filers out most of the fluctuation near the control weights and restore the weight to the Z weight when it exceeds the control threshold which makes the FE weights achieve high resolution without affecting the stability of shock capturing.

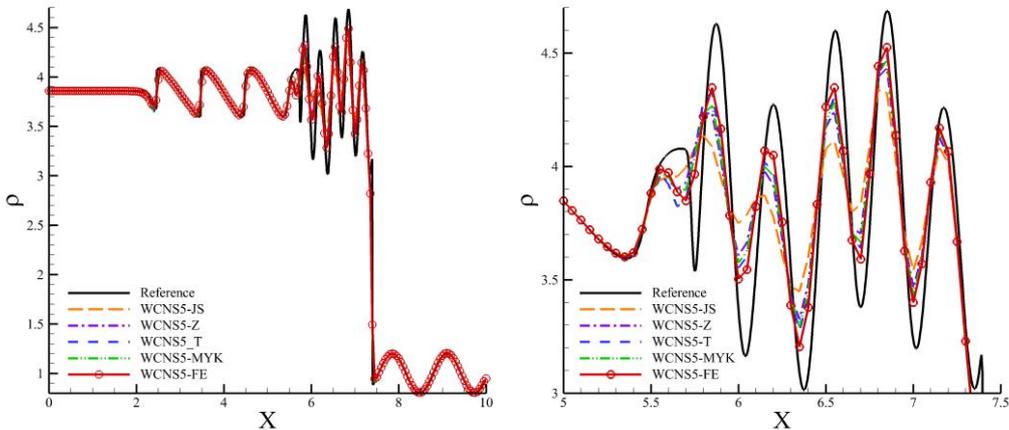

(a) Global profile  (b) Local profile

Fig. 10. Density profiles of the Shu-Osher problem computed by fifth-order weighted compact nonlinear schemes(WCNS) with *N*=201 points presented in (a) Global profile and (b) Local profile

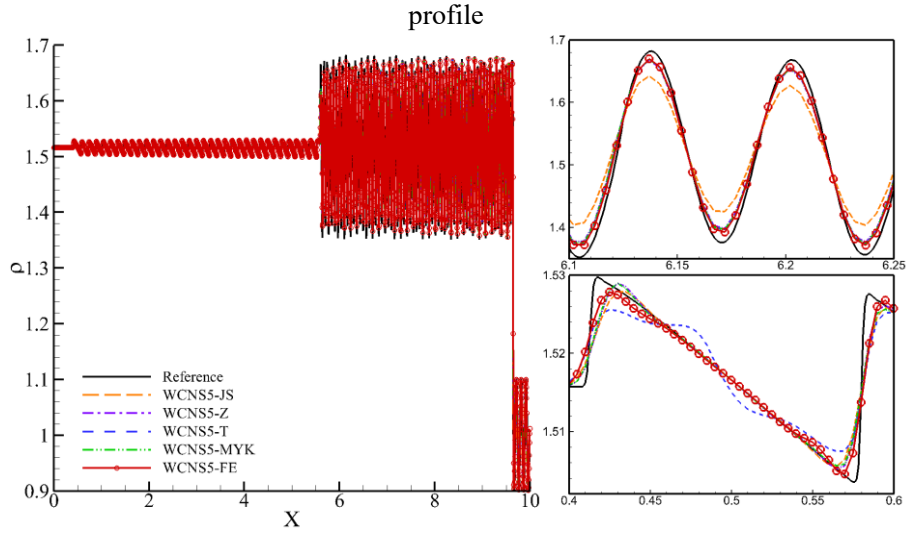

Fig. 11. Density profiles of the Titarev-Toro problem computed by fifth-order weighted compact nonlinear schemes(WCNS) with *N*=2001 points.

**4.3 Double Mach reflection of a strong shock**

This case contains a Mach 10 right-facing shock which is reflected for the wall with angle of $60°$. As the shock moves and reflects on the wall, a complex shock structure with two triple points and a slip line evolves. The computational domain is $[0,4]\times[0,1]$ with the initial condition[29] shown below.

$$(\rho,u,v,p) = \begin{cases} (1.4,0,0,1), & y < \sqrt{3}(x-1/6), \\ (8.0, 7.145, -4.125, 116.5), & else. \end{cases} \quad (37)$$

Inflow and outflow are imposed at left and right boundaries. At the bottom boundary, the inviscid wall boundary conditions are applied on (1/6,4), and post-shock condition is used at the region from [0,1/6]. The values at the top boundary correspond to the exact motion of a Mach 10 shock. Computation marches up to t=0.2 on a $960\times 240$ mesh. Density profile is shown in Fig. 12.

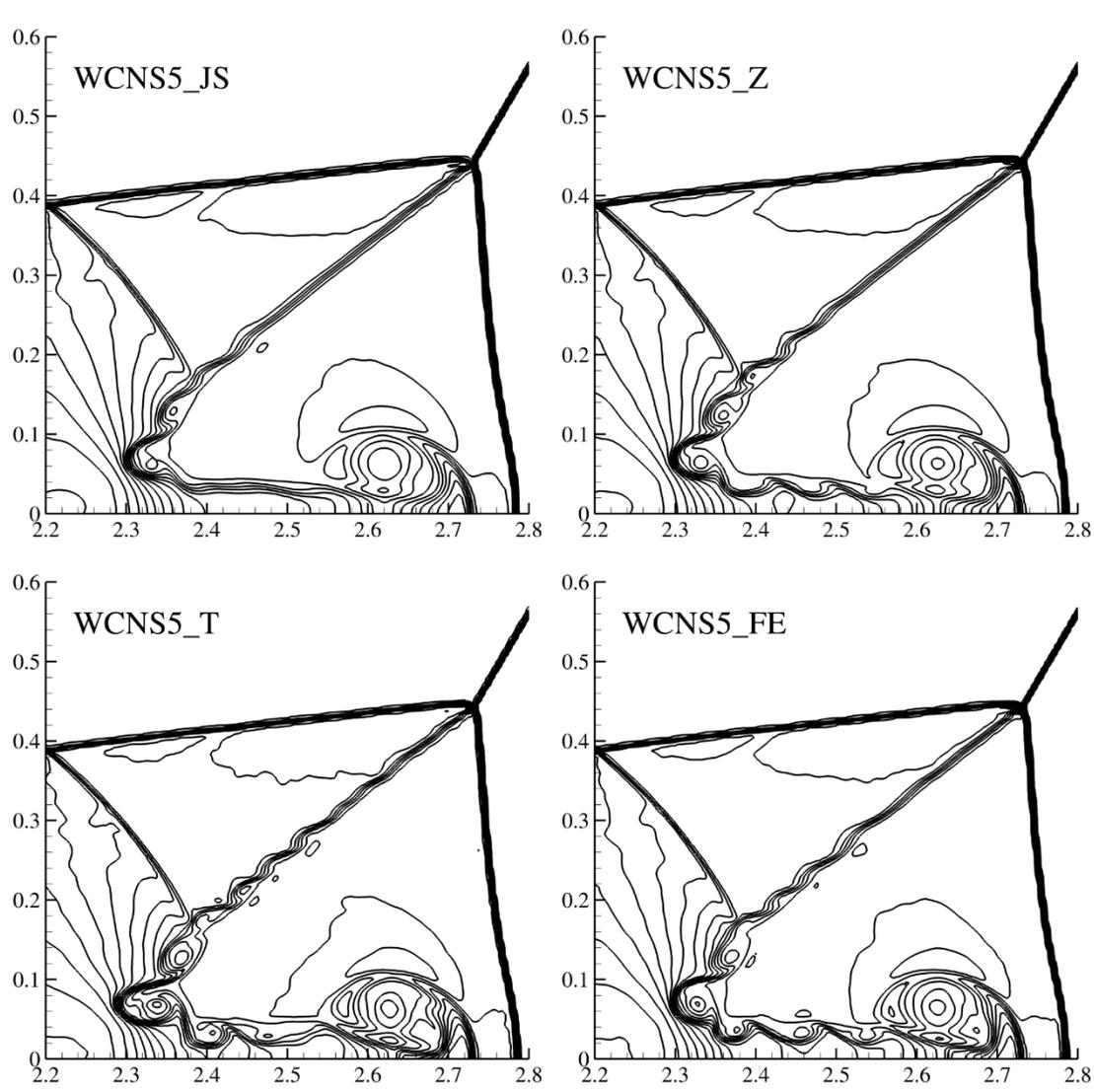

Fig. 12. Density profiles of the Double Mach Reflection problem computed by fifth-order weighted compact nonlinear schemes(WCNS) with $N = 960 \times 240$ points

Due to the Kelvin-Helmholtz instability, the vortices along the discontinuous slip line is a represent for numerical dissipation. Fig.12 shows the consequence containing both slip lines and triple points for four schemes. We can notice that with the same mesh resolution, WCNS5-JS is too dissipative to produce any rolled-up vortices along the slip line and using WCNS5-Z only slightly improves the resolution of vortices. The T weights and FE weights outperform the other schemes as the extra details they capture at slip lines improve vortex resolution. A closer analysis of Fig.12 reveals that FE weights exhibit better performance than WCNS-T, capturing more vortices at the bottom of the figure with less oscillation. These results suggest that our method is physically more robust.

## 4.4 2D Riemann problem

We utilize the WCNS5-FE method for the 2D Riemann problems presented by Schulz-Rinne[30]. The $[0,1]\times[0,1]$ computational domain is partitioned into four quadrants by x=0.8 and y=0.8 lines. Initially constant states are imposed on each of the four quadrants,

$$(\rho,u,v,p) = \begin{cases} (1.5,0,0,1.5), & 0.8 \le x \le 1.0, 0.8 \le y \le 1.0, \\ (0.5323,1.206,0,0.3), & 0.0 \le x \le 0.8, 0.8 \le y \le 1.0, \\ (0.138,1.206,1.206,0.029), & 0.0 \le x \le 0.8, 0.0 \le y \le 0.8, \\ (0.5323,0,1.206,0.3), & 0.8 \le x \le 1.0, 0.0 \le y \le 0.8. \end{cases} \tag{38}$$

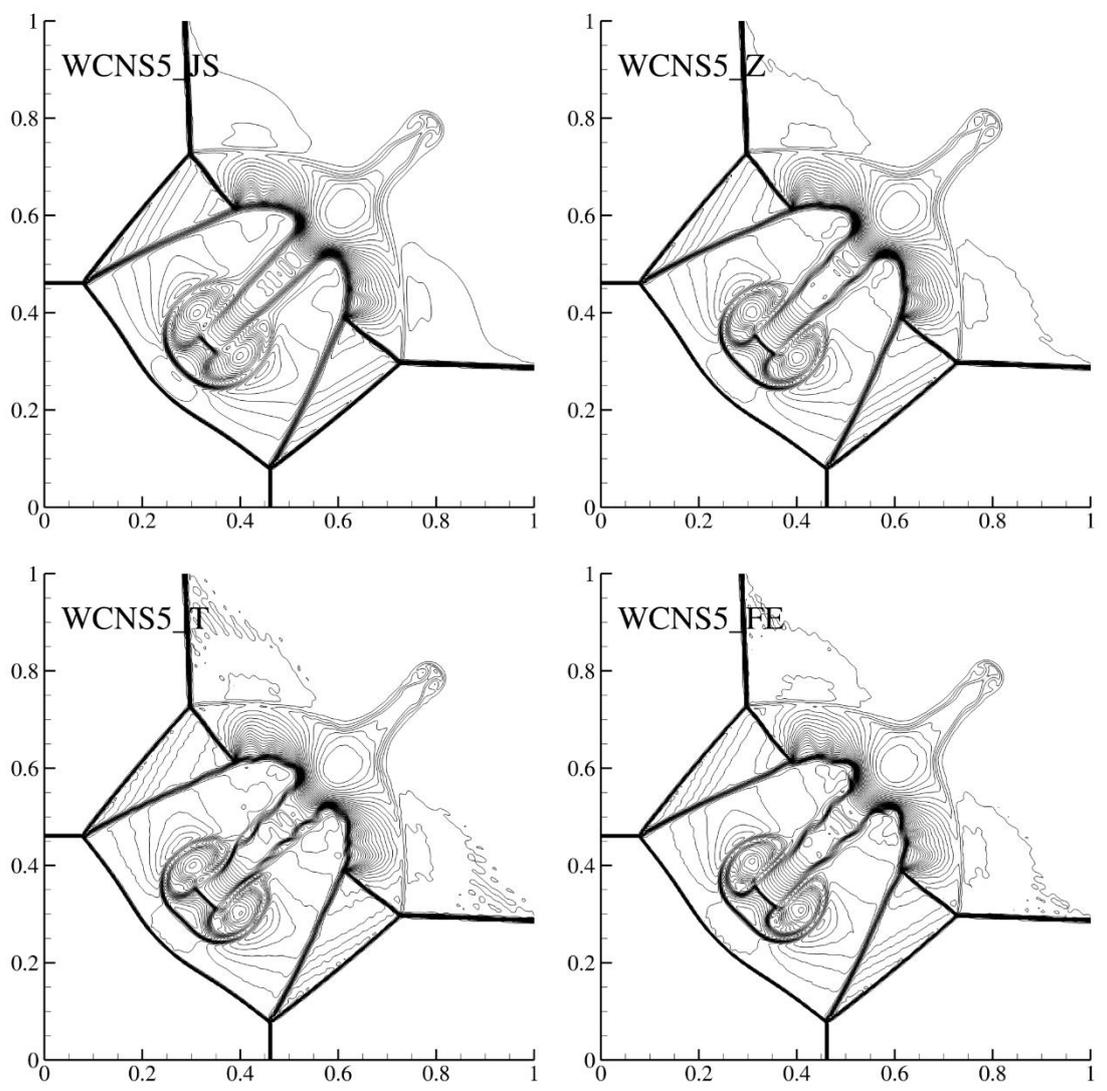

Fig. 13. Density profiles of the 2D Riemann problem computed by fifth-order weighted compact nonlinear schemes(WCNS) with $N = 256\times 256$ points and the final time t=0.8.

Figure 13 shows a comparison of the results for two schemes computed over a grid system of $256 \times 256$. Here, the Mach shock and its symmetric counterpart join and bound the subsonic area. A slip line extending from the branch point into the subsonic area towards the symmetry axis is also present. The Euler equations were solved using shock-capturing schemes, revealing a fine structure which gave an indication of the schemes' performance. However, WCNS5-JS and WCNS5-Z were unable to capture the vortex structures of the slip line due to large dissipation, while WCNS5-FE scheme offer lower dissipation and hence more features can be preserved near the slip line.

# 5 Conclusion

This paper expands the embedded/multistep WENO scheme into the WCNS scheme. Moreover, we present a new scheme, named WCNS-FE, which can overcome the deficiency of WCNS scheme near discontinuities and filter the weight perturbations out. The WCNS-FE scheme combines the stability of the WCNS scheme and the filter property of WCNS-T, providing the benefits of both scheme.

The newly proposed WCNS-FE scheme has several advantages over the original WCNS scheme. Firstly, the scheme has a reasonable distribution of weights as the proportion of ideal weights in the calculation is very small, and the FE weights are obtained through special mapping functions to eliminate perturbations. Meanwhile, due to the embedded process, our scheme is fourth-order in the case of partially discontinuous distribution through this function. Secondly, the continuous and differentiable mapping function ensures the filer property of the scheme is maintained while WCNS-T achieves this through judgement, which is essentially a discontinuous function. Thirdly, in consideration of computation and resolution, we set the reasonable activation of the mapping function based on statistical probability of weights. Finally, the WCNS-FE scheme can be extended to an arbitrary high-order scheme under the WCNS framework.

Results from a series of numerical examples in 1D and 2D scenarios demonstrate that WCNS-FE outperforms other schemes, such as WCNS-JS and WCNS-Z. In fact, WCNS-FE exhibits superior performance even in high wavenumber problems when compared to WCNS-T. Based on the framework of fifth-order WCNS-FE, future work will focus on deriving the higher-order WCNS-FE family schemes and the effects in turbulence simulation.

# Appendix A. PFE weight and Statistical probability analysis

In the Appendix we present the statistical probability of normalized weights during the calculation. Due to the massive computing costs the function needed, we set a switch to determine whether the weight needs to be mapped or not.

$$\tilde{\omega}_i = \begin{cases} \omega_i, & if\ \omega_0 \in [d_0 - c_v, d_0 + c_v].and.\omega_1 \in [d_1 - c_v, d_1 + c_v].and.\omega_2 \in [d_2 - c_v, d_2 + c_v], \\ \omega_i^{mapped}, & else. \end{cases}$$

where, $\tilde{\omega}_i$ is denoted as the partial FE weight denoted as PFE, $i$ range from 0 to 2, and $\omega_i^{mapped}$ is the FE weight. The $c_v$ is the value of the switch which we will discuss the specific value through statistical probability next.

The distribution of the weights after normalization is [0,1] which is discretized into 10000 subdomains with step $\Delta\omega = 10^{-4}$. We make statistics on the probability of the Z weights $\omega_i$, FE weights $\tilde{\omega}_i^{mapped}$ and PFE weights $\tilde{\omega}_i$ distributed in each subdomain. And the region which contains ideal weights is called ideal region here.

## A.1 Shock tube

Both of Sod problem and Lax problem contains shock and low wavenumber area. Fig. A.1 shows the distribution of weights, we can find out that the probability of Z weights distribution in the non-ideal weights region is pretty large and the peak probability is not at the ideal weight, but deviates from the ideal weight. On the contrary, the FE weights reduce the probability of weights distributed in non-ideal region and the peak distributes at the ideal weight.

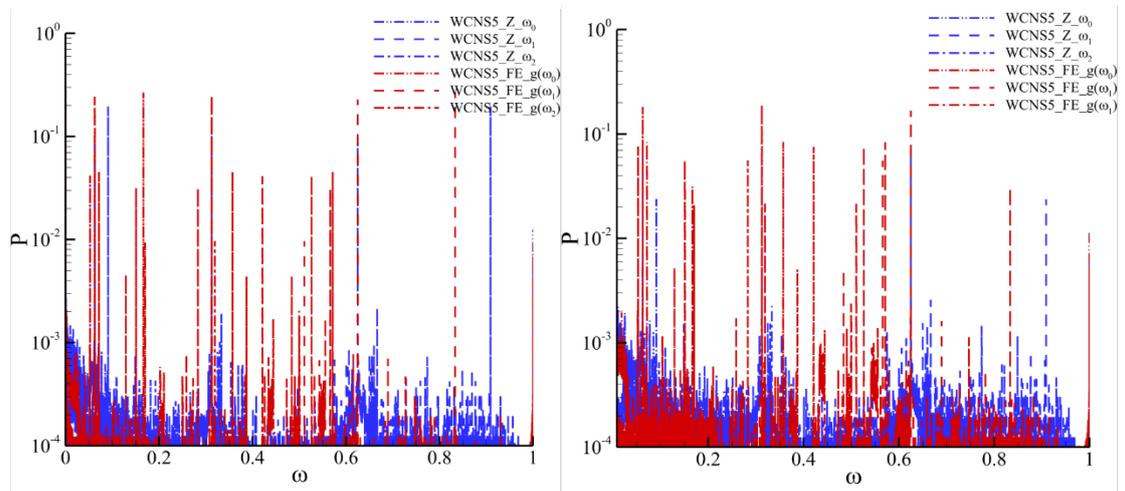

(a)Sod problem　　　　　　　　　　(b)Lax problem

Fig. A.1. Statistical probability of Z and FE weights in shock tube problem. The distribution of the weights after normalization is [0,1] which is discretized into 10000 subdomains with step $\Delta\omega=10^{-4}$.

As shown in Table. A.1, the probability of the ideal weights is presented. Due to the filter property of FE weights, the perturbations can be filtered out and the weights near the ideal region can be mapped to it which can increase the ideal weights ratio. The probability of FE weights distributed in ideal weights region is more than twice large compared to Z weights.

Table. A.1 Statistical probability comparison of Z and FE weights in ideal region for shock tube problem

| Case | | $P(\omega_0)$ | $P(\omega_1)$ | $P(\omega_2)$ |
|---|---|---|---|---|
| Sod | $\omega_i$ | 0.1231 | 0.1093 | 0.1045 |
| | $g(\omega_i)$ | 0.2527 | 0.2498 | 0.2492 |
| Lax | $\omega_i$ | 0.1171 | 0.1043 | 0.1072 |
| | $g(\omega_i)$ | 0.2616 | 0.2575 | 0.2568 |

Table A.1 shows the consequence with or without the switch of mapping function which is called FE weights and PFE weights here. $c_v=0$ means the computation has no switch and FE weights are used. Furthermore, $c_v\neq0$ means the mapping function only be activated when the weights deviating from ideal region within a reasonable range which is determined by $c_v$. And it is PFE weights we mentioned above. The error in the table represents the deviation of $c_v=0$ and $c_v\neq0$ which is calculated by

$$error = (1-\frac{P(\omega_i)_{c_v\neq0}}{P(\omega_i)_{c_v=0}})$$

Table. A.2 Statistical probability deviation of FE weight and PFE weight for shock tube problem

| Case | $c_v$ | $P(\tilde{\omega}_0)$ | error(%) | $P(\tilde{\omega}_1)$ | error(%) | $P(\tilde{\omega}_2)$ | error(%) |
|---|---|---|---|---|---|---|---|
| Sod | 0 | 0.2527 | - | 0.2498 | - | 0.2492 | - |
| | 1.0E-4 | 0.2488 | 1.543 | 0.2456 | 1.681 | 0.2449 | 1.726 |

| | 1.0E-3 | 0.2495 | 1.266 | 0.2428 | 2.802 | 0.2433 | 2.368 |
| | 1.0E-2 | 0.2359 | 6.648 | 0.2183 | 12.61 | 0.2230 | 10.51 |
| | 0 | 0.2616 | - | 0.2575 | - | 0.2568 | - |
| Lax | 1.0E-4 | 0.2604 | 0.459 | 0.2554 | 0.816 | 0.2546 | 0.857 |
| | 1.0E-3 | 0.2615 | 0.382 | 0.2498 | 2.990 | 0.2510 | 2.259 |
| | 1.0E-2 | 0.2556 | 2.294 | 0.2435 | 5.437 | 0.2447 | 4.712 |

## A.2 Modified shock/turbulence interaction

The Shu-Osher problem contains both shock and low wavenumber, and Titarev-Toro problem contains high wavenumber except for the property mentioned above. And Hence, the two problems presented here can cover a lot of cases.

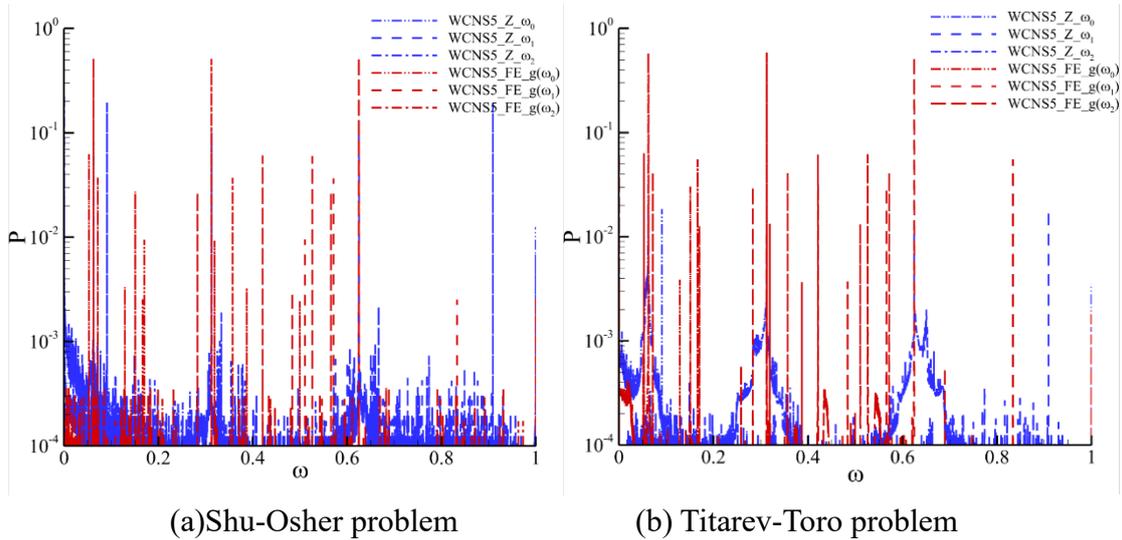

(a) Shu-Osher problem    (b) Titarev-Toro problem

Fig. A.2. Statistical probability of Z and FE weights in modified shock/turbulence interaction. The distribution of the weights after normalization is [0,1] which is discretized into 10000 subdomains with step $\Delta\omega=10^{-4}$.

We can see that the Z weights has severe fluctuation in both problems. And the probability in ideal region is quite small. The FE weights increase the probability in ideal region more than forty times.

Table. A.3 Statistical probability comparison of Z and FE weights in ideal region for Modified shock/turbulence interaction

| Case | | $P(\omega_0)$ | $P(\omega_1)$ | $P(\omega_2)$ |
|---|---|---|---|---|
| Shu Osher | $\omega_i$ | 0.0111 | 0.0031 | 0.0036 |
| | $g(\omega_i)$ | 0.4966 | 0.4903 | 0.4885 |

| | | | | | |
|---|---|---|---|---|---|
| Titarev | $\omega_i$ | 0.0331 | 0.0129 | 0.0137 |
| Toro | $g(\omega_i)$ | 0.5982 | 0.5930 | 0.5915 |

Table. A.4 Statistical probability comparison of Z and FE weights in ideal region for Modified shock/turbulence interaction

| Case | $c_v$ | $P(\tilde{\omega}_0)$ | error(%) | $P(\tilde{\omega}_1)$ | error(%) | $P(\tilde{\omega}_2)$ | error(%) |
|---|---|---|---|---|---|---|---|
| | 0 | 0.4966 | - | 0.4903 | - | 0.4885 | - |
| Shu | 1.0E-4 | 0.4961 | 0.100 | 0.4899 | 0.082 | 0.4882 | 0.061 |
| Osher | 1.0E-3 | 0.4951 | 0.302 | 0.4779 | 2.529 | 0.4771 | 2.333 |
| | 1.0E-2 | 0.3881 | 21.85 | 0.3742 | 23.68 | 0.3734 | 23.56 |
| | 0 | 0.5982 | - | 0.5930 | - | 0.5915 | - |
| Titarev | 1.0E-4 | 0.5962 | 0.334 | 0.5909 | 0.354 | 0.5894 | 0.355 |
| Toro | 1.0E-3 | 0.5953 | 0.485 | 0.5789 | 2.378 | 0.5787 | 2.164 |
| | 1.0E-2 | 0.4818 | 19.46 | 0.4640 | 21.75 | 0.4639 | 21.57 |

In order to decrease computation and keep high resolution, $c_v$ should be as large as possible to reduce the activation of mapping function when it has little impact on consequence. And by observing Table. A.1,A.2,A.3 and A.4, though $c_v$ reducing the effect of probability in ideal region, the time of calculation will increase rapidly and when $c_v = 10^{-3}$ the influence can be controlled within 3%. So $c_v = 10^{-3}$ is the value we recommend here.

## Appendix B. Expend to filtered embedded WENO scheme

In this Appendix, we extend the scheme into WENO and the only change is the coefficient of equation (17). Follow the same process of WCNS-FE, the embedded mapped weights can be introduced as follows:

- If the region is smooth, then $\beta_0 \cong \beta_1 \cong \beta_2$. According to Equation (17), we have

$\omega_0^{original} = \omega_0^{mapped} \to 1/10, \omega_1^{original} = \omega_1^{mapped} \to 6/10, \omega_2^{original} = \omega_2^{mapped} \to 3/10$. And the numerical flux accuracy remains fifth-order.

- If the discontinuity located at $\{x_{i-2}, x_{i-1}\}$, then $\beta_0 \gg \beta_1, \beta_2$. According to

Equation (17), we have $\omega_0^{original} = \omega_0^{mapped} \to 0$, $\omega_1^{original} \to 2/3, \omega_1^{mapped} \to 1/2$ and $\omega_2^{original} \to 1/3, \omega_2^{mapped} \to 1/2$. The numerical flux accuracy remains fourth-order.

- If the discontinuity located at $\{x_{i-1}, x_i\}$, then $\beta_0, \beta_1 \gg \beta_2$. According to Equation (17), we have $\omega_0^{original} = \omega_0^{mapped} \to 0$, $\omega_1^{original} = \omega_1^{mapped} \to 0$ and $\omega_2^{original} = \omega_2^{mapped} \to 1$. The numerical flux accuracy remains third-order.

- If the discontinuity located at $\{x_i, x_{i+1}\}$, then $\beta_1, \beta_2 \gg \beta_0$. According to Equation (17), we have $\omega_0^{original} = \omega_0^{mapped} \to 1$, $\omega_1^{original} = \omega_1^{mapped} \to 0$ and $\omega_2^{original} = \omega_2^{mapped} \to 0$. The numerical flux accuracy remains third-order.

- If the discontinuity located at $\{x_{i+1}, x_{i+2}\}$, then $\beta_2 \gg \beta_0, \beta_1$. According to Equation (17), we have $\omega_0^{original} \to 1/7, \omega_0^{mapped} \to 1/4$, $\omega_1^{original} \to 6/7$, $\omega_1^{mapped} \to 3/4$ and $\omega_2^{original} = \omega_2^{mapped} \to 0$. The numerical flux accuracy remains fourth-order.

The result of embedded interpolation process can be express as Table B.1 briefly.

Table B.1. The mapping correspondence for embedded fifth order WENO schemes

|  | $\omega_0$ | | $\omega_1$ | | $\omega_2$ | |
| --- | --- | --- | --- | --- | --- | --- |
|  | original | mapped | original | mapped | original | mapped |
| Case$_0$ | $\dfrac{1}{10}$ | $\dfrac{1}{10}$ | $\dfrac{6}{10}$ | $\dfrac{6}{10}$ | $\dfrac{3}{10}$ | $\dfrac{3}{10}$ |
| Case$_1$ | - | - | $\dfrac{2}{3}$ | $\dfrac{1}{2}$ | $\dfrac{1}{3}$ | $\dfrac{1}{2}$ |
| Case$_2$ | - | - | - | - | 1 | 1 |
| Case$_3$ | 1 | 1 | - | - | - | - |
| Case$_4$ | $\dfrac{1}{7}$ | $\dfrac{1}{4}$ | $\dfrac{6}{7}$ | $\dfrac{3}{4}$ | - | - |

Similarly, we give the coefficient of embedded WENO mapping function and it is shown in Table B.2 briefly.

Table B.2. The coefficient of mapping function

| i=0 | i=1 | i=2 |
| --- | --- | --- |

|  | $\omega_{0,j}$ | $\psi_{0,j}$ | $\omega_{1,j}$ | $\psi_{1,j}$ | $\omega_{2,j}$ | $\psi_{2,j}$ |
| --- | --- | --- | --- | --- | --- | --- |
| j=0 | $\omega_0$ | 0 | $\omega_1$ | 0 | $\omega_2$ | 0 |
| j=1 | $\dfrac{1}{10}$ | $\dfrac{1}{10}$ | $\dfrac{6}{10}$ | $\dfrac{6}{10}$ | $\dfrac{3}{10}$ | $\dfrac{3}{10}$ |
| j=2 | $\dfrac{1}{4}$ | $\dfrac{1}{7}$ | $\dfrac{1}{2}$ | $\dfrac{2}{3}$ | $\dfrac{1}{2}$ | $\dfrac{1}{3}$ |
| j=3 | $\omega_0$ | 1 | $\dfrac{3}{4}$ | $\dfrac{6}{7}$ | $\omega_2$ | 1 |
| j=4 | - | - | $\omega_1$ | 1 | - | - |